\documentclass{aa}
\usepackage{graphics}
\begin{document}
\thesaurus{08.05.1, 08.23.2, 
           09.04.1, 09.08.1, 09.09.1, 09.11.1,
           11.09.1, 11.13.1  } 

\def\frac{$''$\hspace*{-.1cm}}
\def\deg{$^{\circ}$}
\def\deg{$^{\circ}$\hspace*{-.1cm}}
\def\min{$'$\hspace*{-.1cm}}
\def\h2{H\,{\sc ii}}
\def\hi{H\,{\sc i}}
\def\hb{H$\beta$}
\def\ha{H$\alpha$}
\def\hd{H$\delta$}
\def\heii{He\,{\sc ii}}
\def\hg{H$\gamma$}
\def\sii{[S\,{\sc ii}]}
\def\siii{[S\,{\sc iii}]}
\def\oiii{[O\,{\sc iii}]}
\def\oii{[O\,{\sc ii}]}
\def\hei{He\,{\sc i}}
\def\sm{$M_{\odot}$}
\def\sl{$L_{\odot}$}
\def\ab{$\sim$}
\def\x{$\times$}
\def\sec{s$^{-1}$}
\def\cm2{cm$^{-2}$}
\def\mcube{$^{-3}$}
\def\lam{$\lambda$}

\title{{\it HST} observations of the very young SMC  ``blob''  
   N\,88A\thanks{Based on observations with 
   the NASA/ESA Hubble Space Telescope obtained at the Space Telescope 
   Science Institute, which is operated by the Association of Universities 
   for Research in Astronomy, Inc., under NASA contract 
   NAS\,5-26555.}$^{,}$\,\thanks{Based on observations obtained 
   at the European Southern Observatory, La Silla, Chile.   }
}

\offprints{M. Heydari-Malayeri, heydari@obspm.fr}

\date{Received 27 January 1999 / Accepted 20 April 1999}

\titlerunning{SMC N\,88A}
\authorrunning{Heydari-Malayeri et al.}

\author{M. Heydari-Malayeri\inst{1} \and V. Charmandaris\inst{1} \and
    L. Deharveng\inst{2} \and  M.R. Rosa\inst{3,}\,\thanks
    {Affiliated to the Astrophysics Division, Space Science 
    Department of the European Space Agency.} \and H. Zinnecker\inst{4}
 }

\institute{{\sc demirm}, Observatoire de Paris, 61 Avenue de l'Observatoire, 
F-75014 Paris, France \and
Observatoire de Marseille, 2 Place Le Verrier, 
F-13248 Marseille Cedex 4, France \and
Space Telescope European Coordinating Facility, European 
Southern Observatory, Karl-Schwarzschild-Strasse-2, D-85748 Garching bei 
M\"unchen, Germany \and
Astrophysikalisches Institut Potsdam, An der Sternwarte 16, 
D-14482 Potsdam, Germany 
} 

\maketitle

\begin{abstract}

High-resolution {\it Hubble Space Telescope} images have allowed us
for the first time to resolve the compact SMC ionized ``blob'' N\,88A
(diameter \ab\,3\frac.5 or 1 pc). This very young \h2\, region, which 
is hatching from its natal molecular cloud, is heavily affected by
absorbing dust associated with the cloud. The interstellar reddening
towards N\,88A is on average $A_{V}$\,\ab\,1.5 mag and strikingly
rises to more than 3.5 mag in a narrow dust band crossing the
core of the \h2\, region. Such a high extinction is unprecedented for
an \h2\, region in the metal-poor SMC.  We present the photometry of
some 70 stars lying towards the OB association at the center of which
lies N\,88A.  The exciting star(s) of N\,88A is not detected, due to
the heavy extinction. The chronology of star formation is discussed
for the whole region.
 

\end{abstract}

\section{Introduction}

The {\it Hubble Space Telescope (HST)} offers a unique opportunity for
studying very young massive star formation regions in the outer
galaxies.  The fact that the Small Magellanic Cloud (SMC) is the most
metal-poor galaxy observable with very high angular resolution makes
it an ideal laboratory for investigating star formation in very
distant galaxies reminiscent of those populating the early
Universe. 

Our search for the youngest massive stars in the Magellanic Clouds
started almost two decades ago on the basis of ground-based
observations. This led to the discovery of a distinct and very rare
class of \h2 regions in these galaxies, that we called high-excitation
compact \h2 ``blobs'' (HEBs). So far only five HEBs have been found in
the LMC: N\,159-5, N\,160A1, N\,160A2, N\,83B-1, and N\,11A
(Heydari-Malayeri \& Testor 1982, 1983, 1985, 1986, Heydari-Malayeri
et al.\,1990) and two in the SMC: N\,88A and N\,81 (Testor \& Pakull
1985, Heydari-Malayeri et al.\,1988a).  These objects are expected to
harbor newborn massive stars.  

The first part of our {\it HST} project studying the \h2\, ``blobs''
was devoted to the SMC N\,81 (Heydari-Malayeri et al. \cite{hey99},
hereafter Paper I). The Wide Field Planetary Camera\,2 (WFPC2)
observations allowed us to resolve N\,81 and discover a tight cluster
of newborn massive stars embedded in this nebula of \ab\,10\frac\,
across.  The WFPC2 images also uncovered a striking display of violent
phenomena such as stellar winds, shocks, and ionization fronts,
typical of turbulent starburst regions.

The SMC ``blob'' N\,88A is part of the \h2\, region N\,88 (Henize
\cite{hen}), or DEM\,161 (Davies et al. \cite{dav}) which lies in the Shapley
Wing at \ab\,2\deg.2 (2.4\,kpc) from the main body of the SMC. Other
\h2\, regions lying towards the Wing are from west to east N\,81,
N\,84, N\,89, and N\,90. The HEB nature of N\,88A was first recognized
by Testor and Pakull (\cite{tes}) who used CCD imaging at
\ab\,2\frac\, resolution through \ha, \hb, and \oiii\, filters and IDS
spectroscopy (4\frac\,\x\,4\frac\, aperture) to study the central
component N\,88A.  They found a high excitation object
(\oiii/\hb\,=\,7.8) with an interstellar extinction $A_{V}$\,=\,1.7
mag.  The chemical abundances in N\,88 had previously been determined
by Dufour \& Harlow (\cite{dh}) who, using a 10\frac\,\x\,79\frac\,
slit, found a low-metal content typical of the chemical composition of
the SMC.  CCD photometry and spectroscopy of 10 stars lying around
N\,88A were carried out by Wilcots (\cite{wil}) using the {\sc ctio}
90\,cm telescope.  However, the ground-based observations in general
were unable to reveal the internal morphology and stellar content of
N\,88A. 

The {\it HST} was used for imaging and FOS spectroscopy of 
N\,88A (Kurt et al. \cite{kurt}). These pre-{\sc costar} 
observations, in spite of the effort made in data analysis, 
could not clearly show the internal details of N\,88A.
Garnett et al. (\cite{gar}) revisited the chemical abundances in 
N\,88A on the basis of {\it HST} ultraviolet FOS 
(0\frac.7\,\x\,2\frac.0\, aperture) and ground-based spectra.  

In this paper we present recent {\it HST} observations (GO\,6535) of
N\,88A and its surroundings. In the following sections we elaborate on
the extinction and emission properties of each component and suggest a
plausible scenario on the star formation history of this region.

\begin{figure*}
\caption{A set of {\it HST} views of the SMC nebula N\,88.  
{\bf (a)} A ``true-color'' image of the whole WFPC2 field based on
exposures taken with filters \ha\, (red), \oiii\, (green), and $U$ (blue). 
{\bf (b)} A False color image
of the PC field in \ha.  Field size
\ab\,37\frac\,\x\,37\frac\, (\ab\,11\,\x\,11 pc$^{2}$).  {\bf (c)} A
close-up of the \ha\, image presented in Fig. 1b showing N\,88A and
its neighboring N\,88B. Field size \ab\,12\frac\,\x\,12\frac\,
(\ab\,3.5\,\x\,3.5\,pc$^{2}$). The star lying towards N\,88B is
\#55 (see text). }
\label{rgb}
\end{figure*}

\section{Observations and data reduction}

The observations of N\,88A described in this paper were obtained with
WFPC2 on board the {\it HST} on August 31, 1997 using  
the wide- and narrow-band filters (F300W, F467M, F410M, F547M,
F469N, F487N, F502N, F656N). The observational
techniques, data reduction procedures, and photometry 
are similar to those explained in detail in Paper I.  
A composite image is presented in Fig.\,\ref{rgb}.

The ESO EFOSC Camera at the 3.6\,m telescope was also used on 4 June 1988
for imaging N\,88 through a narrow \ha\, filter (ESO\#507, \lam\,
6565.5\,\AA, $\Delta$\lam\,=\,12\,\AA) with exposure times of 1 and 5
minutes. The detector was a RCA CCD chip (\#11) with 0\frac.36 pixels
and the seeing was \ab\,1\frac.5 ({\sc fwhm}). This \ha\, image is
displayed in Fig.\,\ref{efosc}. Due to its relatively short exposure,
this image displays only the brightest part of the \h2\, emission.

\section{Results}

\subsection {Morphology}

\begin{figure}
\resizebox{\hsize}{!}{\includegraphics{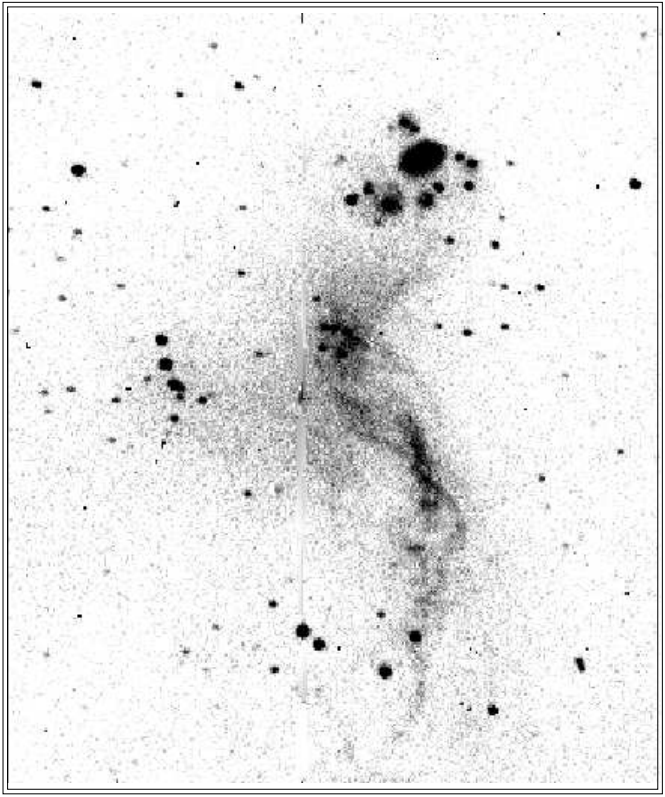}}
\caption{A ground-based \ha\, image of N\,88 obtained 
with EFOSC attached to the ESO 3.6\,m telescope.  Field size
\ab\,1\min.7\,\x\,2\min.0\, (31\,\x\,37 pc$^{2}$).  North is up and
east is left. N\,88A and B appear as the bright ``blob'' at the top of
the image. Note the absence of ionized gas and stars north-west of the
``blob''.}
\label{efosc}
\end{figure}

N\,88 is a relatively large concentration of ionized gas with several
components (Fig.\,\ref{rgb}a). From the central region emanate a
number of fine-structure filaments running southwards over 40\frac\,
(\ab\,10 pc) which can be seen in the ``true-color'' composite image
(Fig.\,\ref{rgb}a).  The larger field of the \ha\, image obtained with
EFOSC (Fig.\,\ref{efosc}) shows a veil of thin filaments curling
southwards over more than 20 pc and brightening at some points.

The main component, N\,88A, is a compact, high excitation \h2\, region
\ab\,3\frac.5\, (\ab\,1\,pc) in diameter surrounded by seven diffuse
\h2\, regions, labelled B to H in Fig.\,\ref{rgb}b.  N\,88A has a
complex morphology. An absorption lane crossing the nebula from north
to south appears as an undulating yellow structure in Fig.\,\ref{rgb}c
(see Sect. 3.2 for more details). West of this structure lies the
brightest part of N\,88A, a small core of diameter \ab\,0\frac.3 (0.08
pc), especially apparent on the \ha\, image (white spot in
Fig.\,\ref{rgb}c; see also Fig.\,\ref{core}).  N\,88A is clearly
ionization-bounded to the north-west since the sharp edge visible in
Fig.\,\ref{rgb} indicates an ionization front in that direction. It
is limited to the south-east by the weaker component B. N\,88B
resembles a hollow sphere -- a shell -- centered on the bright star
\#55 (see Sect. 3.4). N\,88A and N\,88B are clearly in
interaction, as shown by the brightening of the shell between the two
regions. Moreover, we note a high excitation narrow filament 
showing up in the \oiii\, emission north-east of N\,88B 
(Fig.\,\ref{rgb}a). 
The other components are situated farther away from N\,88A.
N\,88\,E-F-G and H appear as more extended, diffuse, and spherical \h2\,
regions.

\begin{figure}
\resizebox{\hsize}{!}{\includegraphics{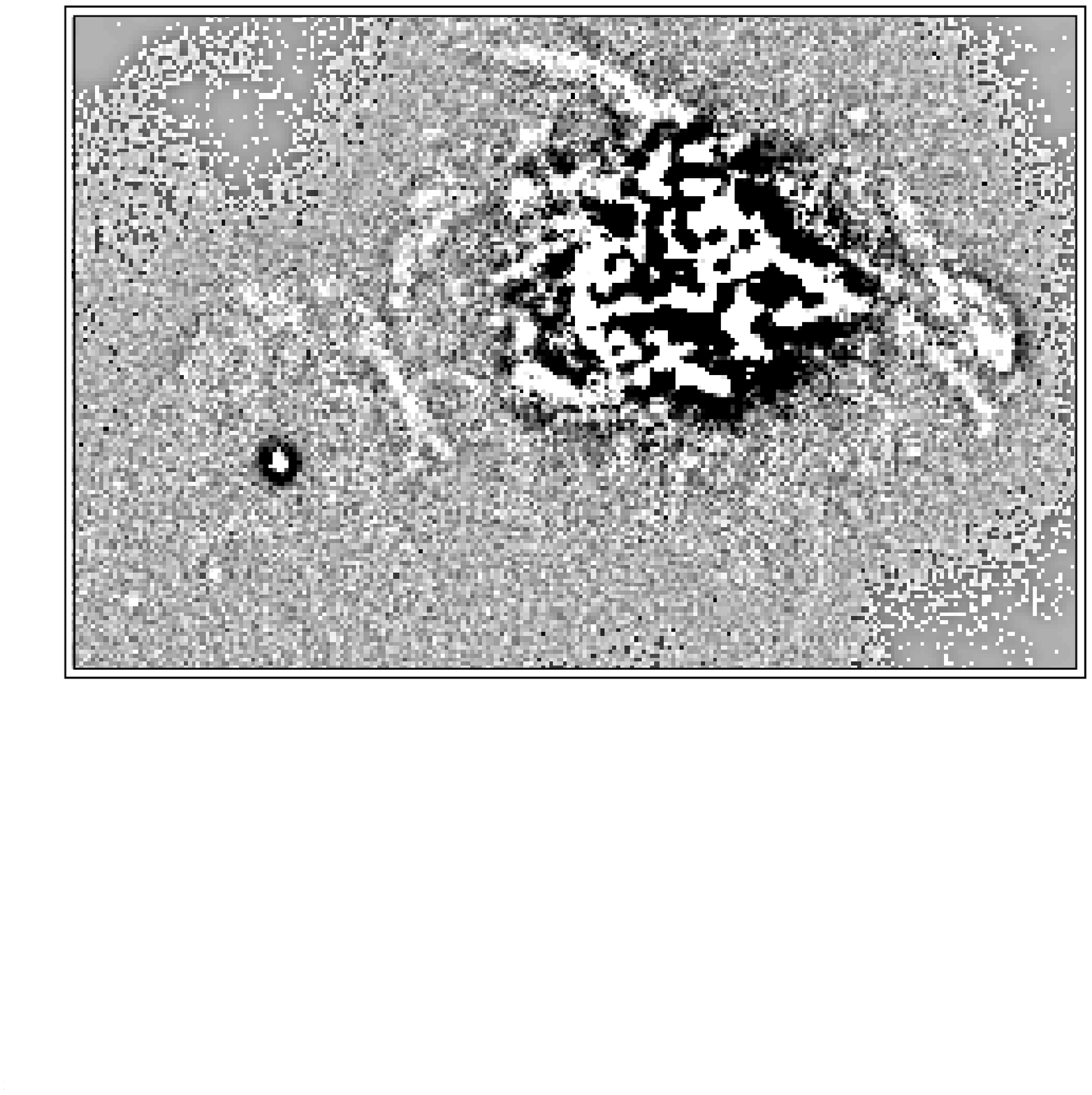}}
\caption{An un-sharp masking \ha\, image of N\,88A \& B 
which highlights the high spatial frequency filamentary patterns.
Field size \ab\,11\frac\,\x\,7\frac.4\, (3.3\,\x\,2.2 pc).  North is
up and east is left. }
\label{vent}
\end{figure}

Several lower excitation arc-shaped features and filaments emerging 
from N\,88A run
outward in the north-east and south-west directions.  These
wind-induced structures are best seen in Fig.\,\ref{vent}, which
presents an un-sharp masked image of N\,88A-B created from \ha. In
this image large-scale structures have been suppressed by the
technique explained in Paper I.  Note also the mottled structure of
the main component A, even in the direction of the absorbing lane,
indicating a very inhomogeneous medium, both for gas and dust, with a
typical cell size of 0\frac.4 (0.1 pc).

\subsection {Nebular reddening}

The Balmer \ha/\hb\, intensity ratio map of N\,88A-B is presented in
Fig.\,\ref{rapport}a. The most striking feature is the presence of a
heavy absorption lane of \ab\,0\frac.7\,\x\,2\frac.3\,
(\ab\,0.2\,\x\,0.7\,pc$^{2}$) in size, running in a north-south
direction, which divides the bright N\,88A into two parts.  The mean
\ha/\hb\, ratio in the lane is 7.10\,$\pm$\,1.42 (rms), corresponding
to $A_{V}$\,=\,2.5 mag if the LMC interstellar reddening law is 
used (Pr\'evot et al. \cite{pre}), 
and reaches values as high as \ab\,10, or
$A_{V}$\,\ab\,3.5 mag.  The mean ratio for component A,
4.81\,$\pm$\,1.46, corresponds to $A_{V}$\,\ab\,1.5 mag. 
The extinction is also high towards component B, where
\ha/\hb\, keeps a relatively uniform value of 4.27\,$\pm$\,0.90
($A_{V}$\,\ab\,1.1 mag).  For comparison, previous lower resolution
spectroscopic observations yielded $A_{V}$\,=\,1.1 mag (Dufour \&
Harlow \cite{dh}, using a 10\frac\, wide slit) and $A_{V}$\,=\,1.7 mag
(Testor \& Pakull \cite{tes}, 4\frac\,\x\,4\frac\, slit).
The \ha/\hb\, map was used to de-redden the \hb\, flux on a 
pixel to pixel basis.

The G component  shows a sharp dividing line in the middle separating it 
into two  distinct halves, one much fainter than the 
other. This feature should be due to absorbing dust.

\subsection{Ionized gas emission}

The total \hb\, flux of component A is F(\hb)\,=\,3.45\,\x\,10$^{-12}$
erg\,\,\sec\,\cm2\, (accurate to \ab\,3\%). Correcting for the reddening
(Sect. 3.2) gives a flux
F$_{0}$(\hb)\,=\,1.85\,\x\,10$^{-11}$ erg\,\,\sec\,\cm2.  The total
flux for both components A and B is
F$_{0}$(\hb)\,=\,1.97\,\x\,10$^{-11}$\,erg\,\,\sec\,\cm2.  Thus,
component B provides less than 10\% of the total \hb\, energy.  A
Lyman continuum flux of $N_{L}$\,=\, 2.10\,\x\,10$^{49}$
photons\,\,\sec\, can be estimated for component A, using a distance
of 66\,kpc, if the \h2\, region is ionization-bounded.  A single O6V
star can account for this ionizing flux (Vacca et al. \cite{vacca},
Schaerer \& de Koter \cite{sch}).  Similarly, the Lyman continuum flux
corresponding to component B is $N_{L}$\,\ab\, 3.5\,\x\,10$^{47}$
photons\,\,\sec. If we take the estimated UV fluxes at face values,
the exciting star of component B should be an early B type star.
However, these should be considered as lower limits, since the \h2\,
regions are not ideally ionization-bounded. 

We find an rms electron density of 2700\,cm\mcube\, for component A
from the total \hb\, flux, a $T_{e}$\,=\,14\,000\,\deg\,K (Garnett et
al. \cite{gar}), and assuming a radius of 0.5\,pc for the object. The
corresponding total ionized mass of N\,88A is \ab\,45\,\sm.

\begin{figure*}
\resizebox{\hsize}{!}{\includegraphics{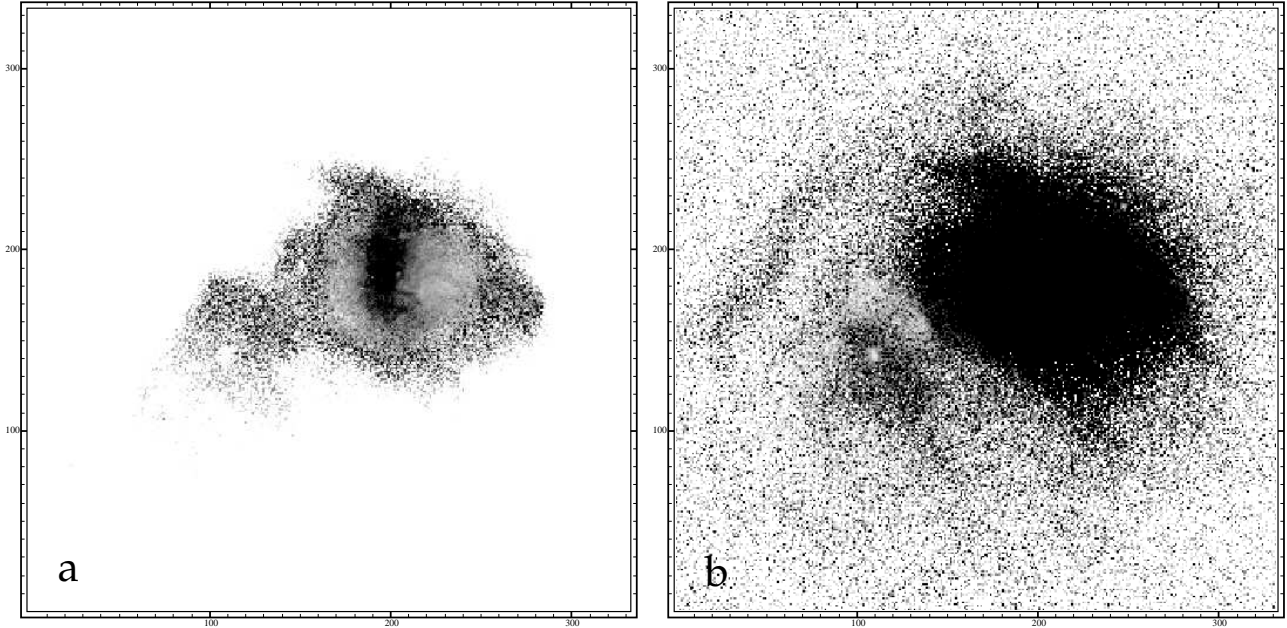}}
\caption{Spatial variation in the extinction and ionization 
across N\,88A. The field size (13\frac.8\,\x\,13\frac.8 or
\ab\,4\,\x\,4 pc) is almost identical to Fig.\,\ref{rgb}c.  {\bf (a)}
Balmer \ha/\hb\, ratio of the inner parts of the nebulae.  Darker
color indicates higher \ha/\hb.  Note the presence of the narrow
absorption lane where the extinction rises to $A_{V}$\,\ab\,3.5 mag.
The mean value of extinction towards component A is about 1.5 mag.
{\bf (b)} The \oiii/\hb\, ratio. The mean value for component A is
\ab\,7 and the ratio rises to as high as 9 at some points. Note the
narrow filament north-east of component B.}
\label{rapport}
\end{figure*}

The \oiii\lam5007/\hb\, intensity map displays an extended high
excitation zone towards N\,88A (Fig.\,\ref{rapport}b), where the ratio
has a mean value of \ab\,7.  The ratio peaks at some
points to values as high as \,9.  Taking
$T_{e}$\,\ab\,14\,000\,\deg\,K and $N_{e}$\,=\,2\,700\,cm\mcube, a
ratio \oiii/\hb\,\ab\,7 indicates an ionic abundance
O$^{++}$/H$^{+}$\,\ab\,9.4\,\x\,10$^{-5}$. Since the mean SMC oxygen
abundance is O/H\,\ab\,10.5\,\x\,10$^{-5}$ (Dufour \cite{duf}), this
means that \ab\,90\% of the oxygen atoms in N\,88A are in the form of
doubly ionized O$^{++}$ ions, in agreement with the result of Garnett
et al. (\cite{gar}).  The \oiii/\hb\, ratio for component B is
comparatively much smaller, with a mean value of \ab\,4.

The high excitation narrow filament emanating from 
component A is clearly visible  in the \oiii/\hb\, map,  
suggesting that the O$^{++}$ ions in the filament may be excited 
by shock collisions.

\subsection{Stars}

The {\it HST} images reveal tens of previously unknown stars towards
the N\,88 complex. Many of them, especially the brightest ones, are
gathered in several small groups often un-resolved in the EFOSC image
(Fig.\,\ref{efosc}).  The photometry obtained for the 79 brightest
stars of the field using the filters wide $U$ (F300W), Str\"omgren $v$
(F410M), Str\"omgren $b$ (F467M), \heii\, (F469N), and Str\"omgren $y$
(F547) is presented in Table\,\ref{phot} which also gives the
coordinates (J2000) of each star.  These stars are identified by their
numbers in Fig.\,\ref{numbers} and Table\,\ref{phot}. 
The capital letters in the
last column of the table identify the associated \h2\, regions.  The
spectral types of the ionizing stars proposed by Wilcots (\cite{wil}), 
as well as their labels, are also listed in the table.  Note that the
present observations show the exciting star of N\,88D to be double
(\#71 \& \#72) and the given type corresponds therefore to both of
them. Table\,\ref{phot} is available in
electronic form at the Centre de Donn\'ees astronomiques de Strasbourg
(via anonymous ftp to cdsarc.u-strasbg.fr or via
http://cdsweb.u-strasbg.fr/Abstract.html). 

\begin{figure}

\resizebox{\hsize}{!}{\includegraphics{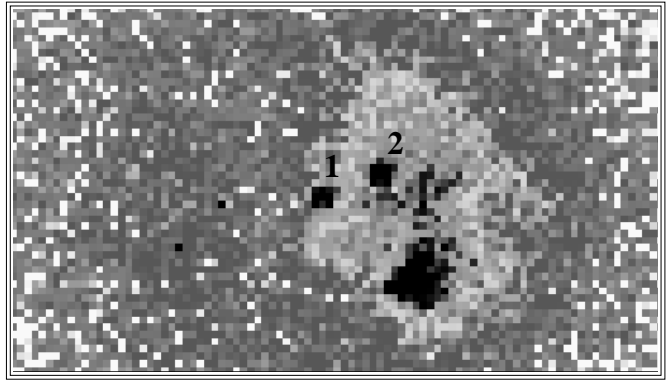}}
\caption{The two stars lying towards the very bright inner 
part of N\,88A (west of the absorption lane) are visible in this
Str\"omgren $y$ image (separated by 0\frac.4, or \ab\,0.1 pc). The
dark spot south of the stars is the brightest core of N\,88A. Field
\ab\,4\frac\,\x\,2\frac\, (\ab\,1.2\,\x\,0.6 pc$^{2}$). Same
orientation as all other images.}
\label{core}
\end{figure}

While the exciting stars of the fainter \h2\, regions are easily
identified on the true-color image, a remarkable point is the absence
of prominent stars towards the main component A. Nevertheless, we
detect two faint stars embedded in the core of N\,88A east of the
absorption lane (Fig.\,\ref{core}).  These are stars \#1 and \#2 with
$y=18.2$ and 18.3 mag and colors $b-y=+0.9$ and +0.6 mag respectively.
It should however be underlined that these magnitudes are very
uncertain since the stars lie in a very bright area where nebular
subtraction is not straightforward.  A third fainter star
($y$\,\ab\,20 mag), not visible in Fig.\,\ref{core}, is marginally
detected just to the east of the bright \ha\, core of N\,88A. 
Its position suggests it as being a good candidate for an 
exciting star.  

Star \#55, situated towards the center of component B, has $y=16.57$
mag and is one of the brightest in the field.  It has a highly
elongated PSF profile which is due to multiplicity (at least three
components are resolved). This may be the star ``$s1$'' detected by
Wilcots (\cite{wil}) relatively close to the brightest part of
N\,88. Its spectrum shows strong
\hei\,\lam\,4471\,\AA\, and \heii\,\lam\,4686, but weak  
\heii\,\lam\,4541\,\AA\, indicative of an O9~V star. 
If this spectrum  belongs actually to \#55, it should correspond 
to the brightest component of this system. 

The color-magnitude diagram for the brightest stars of the sample 
(Fig.\,\ref{cm}) shows a main sequence with the bulk 
of the stars centered on Str\"omgren  
colors $b-y=-0.10$ and $v-b=-0.20$ mag, typical of massive OB stars 
(Relyea \& Kurucz \cite{rk}). These colors are equivalent to 
a Johnson $B-V$\,=\,--0.30 (Turner \cite{tur}), which indicates a 
negligible reddening (Conti et al. \cite{conti}). This result 
is due to the fact that the main sequence is overwhelmingly dominated by 
stars lying outside the N\,88 complex and  means that the 
areas situated north-east, east, south, and south-west of N\,88 are not 
affected by dust.  

However, taking a sub-sample made up of all the exciting stars of the
N\,88 \h2\, regions (excluding stars \#1 \& \#2), we find the
Str\"omgren colors $b-y=0.02$ and $v-b=-0.17$ which indicate
$B-V$\,=\,--0.21 corresponding to a visual extinction of
$A_{V}$\,=\,0.3 mag. This clearly confirms that the N\,88 complex is the
most reddened part of this region of the SMC.  Of particular interest
are stars \#60 and \#61 situated immediately north-west of N\,88A
(Fig.\,\ref{rgb}b).  Assuming that these two stars are of O type,
their colors suggest an extinction of $A_{V}$\,$>$\,1 mag.  This
result has implications for the location of the molecular cloud
(Sect. 4.4).

\begin{figure*}
\caption{A finding chart indicating the brightest stars detected 
towards N\,88, based on the WFPC2 \heii\, image.  The field of view is
the same as in Fig.\,\ref{rgb}a.  North is up and east is left. The
photometry of those stars is presented in Table\,\ref{phot}.}
\label{numbers}
\end{figure*}

The brightest stars of the sample are \#39, \#12, \#19, and \#6. The
first two are blue stars and the latter ones are red.  The red
population showing up in the color-magnitude diagram represents a
collection of evolved field stars as well as young massive ones
contaminated by nebulosity/dust.  For instance, it is noteworthy that
the very red stars \#76, and \#74 are not associated with a
nebulosity, and this suggests that they should be evolved field
stars.

In the particular case of stars \#1 and \#2 lying inside N\,88A, in
spite of their red colors, they may be young blue stars suffering from
heavy extinction.  Assuming that star \#1 has an O9\,V spectrum with
$M_{V}$\,=\,--4.4 mag (Vacca et al. \cite{vacca}), and considering
that the distance modulus of SMC is 19 mag, then an extinction of
$A_{V}$\,=\,3.4 mag is necessary to make it appear as faint as
$y$\,\ab\,18 mag.  Similarly, in order for a star of spectral type
O6\,V (Sect. 3.3) to have an $y$\,\ab\,20 mag, we need an
$A_{V}$\,\ab\,6 mag.  Thus, the main exciting star(s) of N\,88A should
remain hidden in the optical due to the presence of dust by an
extinction of at least 6 mag.

\begin{figure*}
\rotatebox{-90}{\resizebox{10.cm}{!}{\includegraphics{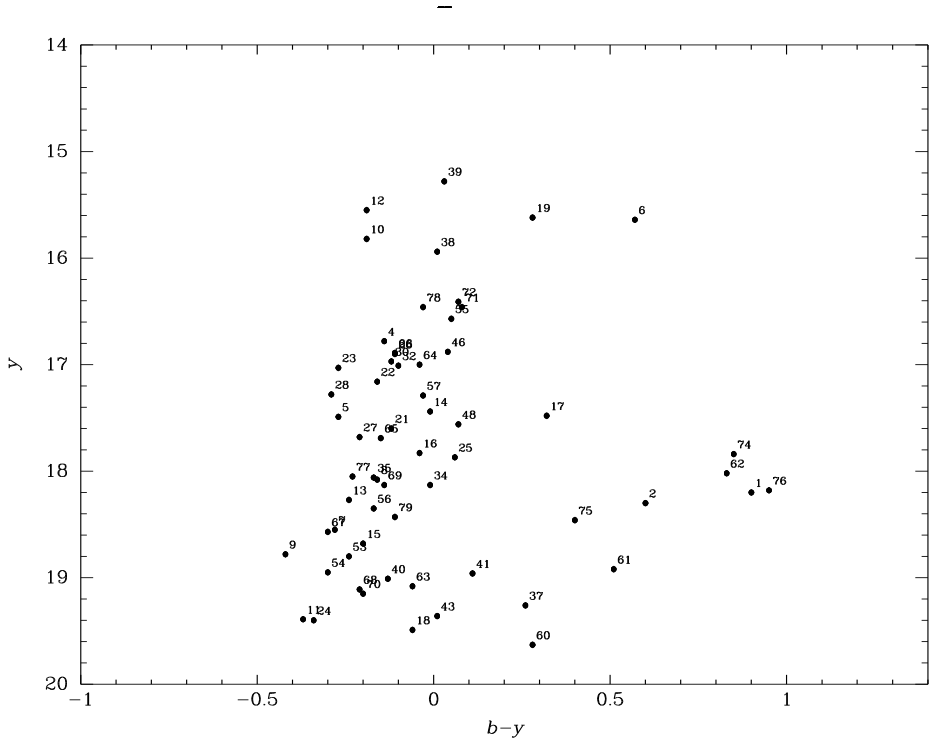}}}
\caption{Color-magnitude diagram,  $y$ versus $b$\,--\,$y$, 
of the brightest stars revealed towards the SMC compact \h2\, region 
N\,88. Numbers refer to Fig.\,\ref{numbers} and Table\,\ref{phot}. } 
\label{cm}
\end{figure*}

\section{Discussion and concluding remarks}

\subsection{Comparison with N\,81}

The most striking feature of N\,88A is its lack of prominent stars,
even at the WFPC2 resolution. This indicates a young age and is
supported by other observational findings about N\,88A: its
compactness, its high density, and its exceptionally high
extinction. These facts considered together suggest that N\,88A is
just hatching from its natal molecular cloud. Stars \#1 and \#2 are
probably among the exciting sources of N\,88A. Other exciting stars
may still be embedded in the densest part of the nebula, such as he
bright spot highlighted in Fig.\,\ref{rgb}c and Fig.\,\ref{core}, and
remain invisible due to the high dust content.  Compared with N\,81,
N\,88A is probably younger since N\,81 is more extended, less dense,
and exhibits several of its exciting stars (Paper I).  Although N\,81
and N\,88A are both very young, the present observations underline
their difference. Apart from the exciting stars aspect, N\,88A is
surrounded by several diffuse \h2\, regions. In contrast, N\,81 is an
isolated object. These facts point to the diversity of star forming
regions belonging to the same chemical environment and also to the
necessity of observing each case in detail.  

On the other hand, the whole N\,88 region is very reminiscent of the
LMC N\,59 region studied by Armand et al. (\cite{arm}). N\,88 and
N\,59 contain several individual \h2\, regions, in various
evolutionary states. They range from compact, bright and young
components with a lot of dust hiding the exciting stars (N\,88A and
N\,59A) to diffuse, spherical and evolved regions (N\,88E-F-G-H and
N\,59C), and also to shell components (N\,88B and N\,59B which
contains a supernova remnant). Similarly, both regions display a
filamentary structure which results from the interaction of the strong
stellar winds emitted by the massive stars with the surrounding
medium, as well as small scale brightness variations pointing to a
very inhomogeneous distribution of matter or dust inside or around
these objects.

\subsection{Associated neutral material}

CO emission from the molecular cloud associated with N\,88A was
observed by Israel et al. (\cite{isco}) using the ESO-SEST 15\,m
telescope.  They detected the $^{12}$CO(1--0) emission with a
brightness temperature of 750 mK, a width of 2.5 km\,\sec\, and a
radial velocity of V$_{LSR}$\,=\,147.8 km\,\sec. The molecular cloud
is much brighter than the one associated with N\,81 (Paper I) 
and ranks among the few sources in the SMC
detected in $^{13}$CO(1--0) (Israel et al. \cite{isco}). Rubio et
al. (\cite{rub}) mapped the molecular cloud in $^{12}$CO(1--0) and
$^{12}$CO(2--1) transitions using the SEST telescope with respective
spatial resolutions of 43\frac\, (\ab\,13 pc) and 22\frac\, (\ab\,7
pc).  It turns out that the cloud is in fact relatively small,
\ab\,1\min\, (18 pc) in size in the east-west direction and 
slightly smaller in north-south. More recently, Rubio et al. (private
communication) have detected molecular transitions $^{12}$CO(3--2),
CS(2--1), CS(3--2), HCO$^{+}$(1--0) which probably originate from the
hot and dense core of the cloud.  Molecular hydrogen emission also has
been detected towards N\,88A (Israel \& Koornneef \cite{ik88}).

The SMC is known to have an overall complex structure with several
overlapping neutral hydrogen layers (McGee \& Newton \cite{McGee}).
We used the recent observations by Stanimirovic et al. (\cite{stan}),
with a resolution of 98\frac\, (30\,pc), to examine the \hi\, emission
towards N\,88.  The \hi\, spectrum profile has two main emission peaks
at \ab\,150 and \ab\,175 km\,\sec.  The column density corresponding
to their sum is 3.12\,\x\,10$^{21}$ atoms\, cm$^{-2}$, slightly
smaller than that corresponding to N\,81 (Paper I).
It seems that the molecular cloud is correlated
with the smaller velocity \hi\, component.

\subsection{Extinction}

N\,88 was detected as a very bright IRAS source (Schwering \& Israel
\cite{schwering}).  The fact that near infrared photometry of N\,88A,
at $J$, $H$, $K$, and $L'$ bands obtained using a 10\frac\, aperture
(Israel \& Koornneef \cite{ik91}), is consistent with the IRAS
spectrum (12, 25, 60, and 100\,$\mu$m) suggested that the IR emission
arises mostly from the compact object in the aperture.  Moreover,
these authors found a quite red $K$\,--\,$L'$ color of more than 2 mag
indicating the presence of hot dust. 

Our {\it HST} images for the first time show the heavy concentration
of absorbing dust towards the inner parts of the \h2\, region.  More
strikingly, the extinction rises to as high as $A_{V}$\,\ab\,3.5 mag
in a narrow band towards the bright core of the nebula. This high
absorption is quite unexpected for a metal-poor galaxy like the SMC.
In fact N\,88A holds the record of extinction among ionized
nebulosities in the SMC.  The correlation between the zones of the
high excitation and high extinction is an argument in favor of the
physical association of the dust with hot gas. 

It is important to know the properties of this dust. Roche et al. 
(\cite{roche}) studied 8--13\,$\mu$m spectra of N\,88A and found 
a featureless continuum without any evidence of dust signatures 
attributed to silicate grains. This led them to the conclusion that 
the dust is likely composed of carbon grains. Further progress 
in this area requires appropriate IR observations using 
the highest spatial resolutions. 

\subsection{Star formation}

The N\,88 nebular complex results from a small starburst which occurred
recently in the Wing of the SMC. While the main stars creating N\,88A
are not visible, the other members of the starburst show up in the
{\it HST} images (Fig.\,\ref{rgb}).  The stars exciting the diffuse
\h2\, regions (C to H) were formed in the outer, less dense parts of
the molecular cloud, whereas the compact, very dusty N\,88A is
associated with the core of this small molecular cloud (Sect. 4.2).
The cloud must be to the north-east of N\,88A, as indicated by the
ionization front detected in that direction (Sect. 3.1) and also by
the fact that stars \#60 and \#61 situated north of the front are
heavily affected by extinction (Sect. 3.4).  This is further supported
by ground-based higher exposure images showing a large front
north-west of N\,88A beyond which no stars are visible
(Fig.\,\ref{efosc}; also Testor \& Pakull \cite{tes}). 

The case of component B is interesting. Although it is, like N\,88A,
apparently related to the core of the molecular cloud, it seems more
evolved. In fact N\,88B has a significantly lower density and less
dust, and reveals its exciting star (\#55). It can be considered that
N\,88A has resulted from sequential star formation, that is the
collapse of the shock/ionization front layer created by stars \#55. If
so, we are dealing with two successive generations of stars formed in
the core of the molecular cloud.

The stars situated towards N\,88 are also known as HW\,81 following Hodge
\& Wright (\cite{hw}) who surveyed the SMC in search of OB
associations. The present observations reveal the fainter members of
this association.  The {\it HST} images also resolve another
association in the direction of N\,88. Lying \ab\,50\frac\, (15 pc)
south-east of N\,88A, at the lower-left corner of Fig.\,\ref{rgb}a,
HW\,82 (Hodge \& Wright \cite{hw}) is composed of a dozen stars
several of which are tightly packed multiple ones. HW\,82 is not
associated with ionized gas, and a relatively large number of its
stars are red. At present we do not know whether the red and blue
stars are co-spatial members of the same cluster.  Nevertheless, the
facts that the ionized gas is already dispersed from there and that no
significant amount of dust is detected (Sect. 3.4) suggest that HW\,82
represents an older burst of star formation in the Wing.  This is
confirmed by the larger field of EFOSC \ha\, image (Fig.\,\ref{efosc})
which shows no \h2\, regions south of the {\it HST} field of view.
Star formation must have therefore proceeded from south to north and
N\,88 is the most recent site of star formation in this part of the SMC.

A noteworthy aspect of the stellar population towards N\,88 is the
presence of several tight clusters or multiple systems uncovered by
the present observations. For example, the exciting star of N\,88B
(\#55) is a multiple system of at least three components. There are
also at least two stars hidden inside N\,88A, while both N\,88C and D
are excited by two blue stars of comparable brightness.  Note also the
tight cluster in HW\,82 composed of stars \#9, \#10, \#11, \#12 , and
\#13.  These cases present new pieces of evidence in support of
collective formation of massive stars in the SMC (see Paper I for a
brief discussion). 

An intriguing, though unanswered question, is related to the origin of
the large-scale filamentary veil visible in the EFOSC image.  Our
true-color image shows that filaments originating from north-east of N\,88
run towards the anonymous blue cluster in the south (stars \#16, \#17,
\#21, \#22, \#23, and \#24).  However, the veil significantly
brightens south of that cluster and bends to the south-east.  In
consequence, the association of the veil with the N\,88 region is not
established. It is possible that this filamentary structure is linked
to the neighboring huge bubble nebula DEM\,167 (Davies et
al. \cite{dav}).

\begin{acknowledgements}
We are grateful to an anonymous referee for his careful reading of the 
manuscript and comments that contributed to substantially improve 
the paper. VC would like to acknowledge the financial support from a 
Marie Curie fellowship (TMR grant ERBFMBICT960967).
\end{acknowledgements}

{}

\begin{table*}[htbp]
\caption[]{Photometry of the brightest stars towards N\,88}
\label{phot}
\begin{tabular}{ccccccccc}   
\hline
Star &  $\alpha$  &   $\delta$    & wide $U$ &  $v$    &  $b$    & \heii   &  $y$ & Notes$\dag$   \\
     & (J2000) & (J2000)   & (F300W)  & (F410M) & (F467M) & (F469N) & (F547M) & \\
\hline
1  &  1:24:08.10  &  -73:09:03.98  &  --       & --      &  19.1   &   --    & 18.2 &   \\
2  &  1:24:08.02  &  -73:09:03.84  &  --       & --      &  18.9   &   --    & 18.3 &  \\	
4  &  1:24:25.42  &  -73:10:25.20  & 14.83 & 16.35 &  16.64 & 16.70 & 16.78 &\\	
5  &  1:24:27.12  &  -73:10:24.20  & 15.56 & 17.10 &  17.22 & 17.14 & 17.49 &  \\	
6  &  1:24:27.58  &  -73:10:19.97  & 16.82 &  16.94 & 16.21 & 16.27 & 15.64 &  \\	
7  &  1:24:27.39  &  -73:10:19.50  & 16.84 &  18.21 & 18.27 & 18.42 & 18.55 &  \\	
8  &  1:23:57.75  &  -73:10:17.77  & 16.51 &  17.74 & 17.92 & 17.95 & 18.08 &  \\	
9  &  1:24:28.00  &  -73:10:15.11  & 16.97 &  18.19 & 18.36 & 18.59 & 18.78 &  \\	
10 &  1:24:28.29  &  -73:10:13.63  & 13.54 &  15.34 & 15.63 & 15.73 & 15.82 &  \\	
11 &  1:24:27.83  &  -73:10:13.06  & 17.91 &  18.92 & 19.02 & 19.30 & 19.39 &  \\	
12 &  1:24:28.23  &  -73:10:12.98  & 13.39 &  15.14 & 15.36 & 15.42 & 15.55 &  \\	
13 &  1:24:28.09  &  -73:10:12.39  & 16.42 &  17.86 & 18.03 & 17.92 & 18.27 &  \\	
14 &  1:24:21.08  &  -73:10:10.12  & 15.86 &  17.33 & 17.43 & 17.47 & 17.44 &  \\	
15 &  1:24:20.38  &  -73:10:10.01  & 17.32 &  18.36 & 18.48 & 18.54 & 18.68 &  \\
16 &  1:24:14.68  &  -73:10:10.16  & 16.19 &  17.62 & 17.79 & 17.89 & 17.83 &  \\	
17 &  1:24:16.27  &  -73:10:08.28  & 18.52 &  18.28 & 17.80 & 17.76 & 17.48 &  \\	
18 &  1:24:26.37  &  -73:10:06.22  & 18.91 &  19.43 & 19.43 & 19.80 & 19.49 &  \\	
19 &  1:24:28.52  &  -73:10:05.18  & 16.25 &  16.19 & 15.90 & 15.98 & 15.62 &  \\		
21 &  1:24:14.42  &  -73:10:03.10  & 15.77 &  17.34 & 17.48 & 17.49 & 17.60 &  \\	
22 &  1:24:15.30  &  -73:10:01.91  & 15.29 &  16.89 & 17.00 & 17.10 & 17.16 &  \\	
23 &  1:24:16.08  &  -73:10:01.38  & 14.96 &  16.60 & 16.76 & 16.93 & 17.03 &  \\	
24 &  1:24:13.82  &  -73:10:01.50  & 17.72 &  18.98 & 19.06 & 19.30 & 19.40 &  \\	
25 &  1:24:07.39  &  -73:10:01.03  & 16.47 &  17.71 & 17.93 & 17.93 & 17.87 &  \\
26 &  1:23:59.63  &  -73:09:48.58  & 15.06 &  16.63 & 16.79 & 16.77 & 16.90 &  \\	
27 &  1:24:02.37  &  -73:09:48.20  & 15.72 &  17.35 & 17.47 & 17.57 & 17.68 &  \\	
28 &  1:24:22.40  &  -73:09:43.50  & 15.40 &  16.82 & 16.99 & 17.19 & 17.28 &  \\		
30 &  1:24:03.05  &  -73:09:34.47  & 15.30 &  16.65 & 16.85 & 16.82 & 16.97 &  \\		
32 &  1:24:06.48  &  -73:09:32.96  & 15.20 &  16.80 & 16.91 & 16.94 & 17.01 &  \\		
34 &  1:24:11.99  &  -73:09:26.63  & 16.74 &  18.11 & 18.12 & 18.14 & 18.13 &  \\	
35 &  1:24:02.55  &  -73:09:25.42  & 16.70 &  17.87 & 17.89 & 17.98 & 18.06 &  \\		
37 &  1:24:07.62  &  -73:09:21.20  & 18.97 &  19.50 & 19.52 & 19.71 & 19.26 &  \\	
38 &  1:24:11.09  &  -73:09:20.59  & 14.20 &  15.72 & 15.95 & 15.89 & 15.94 & H (B0\,III, s3)  \\
39 &  1:24:13.94  &  -73:09:19.47  & 13.39 &  14.98 & 15.31 & 15.30 & 15.28 &  \\	
40 &  1:24:27.56  &  -73:09:17.94  & 17.72 &  18.89 & 18.88 & 18.99 & 19.01 &  \\	
41 &  1:24:08.69  &  -73:09:19.24  & 18.15 &  18.53 & 19.07 & 19.06 & 18.96 &  \\	
43 &  1:24:07.69  &  -73:09:17.90  & 18.01 &  18.92 & 19.37 & 19.14 & 19.36 &  \\		
46 &  1:24:07.21  &  -73:09:15.60  & 15.25 &  16.63 & 16.92 & 16.81 & 16.88 &  C\\		
48 &  1:24:07.37  &  -73:09:14.63  & 15.88 &  17.32 & 17.63 & 17.54 & 17.56 &  C\\		
53 &  1:24:20.76  &  -73:09:08.05  & 17.75 &  18.41 & 18.57 & 18.63 & 18.81 &  \\	
54 &  1:24:15.05  &  -73:09:06.80  & 17.39 &  18.21 & 18.65 & 18.68 & 18.95 &  \\	
55 &  1:24:09.09  &  -73:09:06.05  & 15.21 &  16.57 & 16.62 & 16.62 & 16.57 &  B\\	
56 &  1:24:14.64  &  -73:09:05.42  & 16.91 &  18.08 & 18.18 & 18.30 & 18.35 &  \\	
57 &  1:24:05.66  &  -73:09:05.06  & 15.50 &  16.97 & 17.26 & 17.26 & 17.29 &  F\\		
60 &  1:24:07.65  &  -73:09:02.22  & 18.51 & 19.83 & 19.91 & 19.33 & 19.63 &  \\	
61 &  1:24:06.91  &  -73:09:01.67  & 17.87 & 19.04 & 19.43 & 18.85 & 18.92 &  \\	
62 &  1:24:09.05  &  -73:08:55.90  &   --- & 19.70 & 18.85 & 18.54 & 18.02 &  \\	
63 &  1:24:10.88  &  -73:08:55.38  & 17.78 & 18.86 & 19.02 & 18.79 & 19.08 &  \\	
64 &  1:24:09.88  &  -73:08:53.65  & 15.49 & 16.86 & 16.96 & 16.95 & 17.00 & G (O9.5\,I, \#4)  \\	
65 &  1:24:26.97  &  -73:08:44.10  & 15.74 & 17.28 & 17.54 & 17.46 & 17.69 &  \\	
66 &  1:24:22.46  &  -73:08:41.42  & 15.29 & 16.52 & 16.78 & 16.76 & 16.89 &  \\	
67 &  1:24:13.95  &  -73:08:36.73  & 17.21 & 18.29 & 18.27 & 18.51 & 18.57 &  \\	
68 &  1:24:17.58  &  -73:08:29.33  & 18.48 & 19.37 & 18.90 & 19.58 & 19.11 &  \\	
69 &  1:24:24.35  &  -73:08:28.14  & 16.26 & 17.75 & 17.99 & 18.03 & 18.13 &  \\	
70 &  1:24:16.32  &  -73:08:14.91  & 17.74 & 18.53 & 18.95 & 18.66 & 19.15 &  \\	
71 &  1:24:04.97  &  -73:09:15.11  & 15.05 & 16.55 & 16.54 & 16.46 & 16.46 & D (B0\,III, \#6) \\ 
72 &  1:24:04.97  &  -73:09:14.38  & 14.97 & 16.42 & 16.48 & 16.31 & 16.41 & D \\
73 &  1:24:04.80  &  -73:09:07.55  &  --   &  --   & --    &  --   & --    & E (B0\,V, \#5)\\
74 &  1:24:05.23  &  -73:10:03.39  & 20.42 & 19.96 & 18.69 & 18.52 & 17.84 & \\

\end{tabular}
\end{table*}

\begin{table*}[htbp]
{\bf Table 1.} (continued)
\begin{flushleft}
\begin{tabular}{ccccccccc}   
\hline
Star &  $\alpha$  &   $\delta$    & wide $U$ &  $v$    &  $b$    & \heii   &  $y$ & Notes$\dag$  \\
     &  (J2000)   &   (J2000)     & (F300W)  & (F410M) & (F467M) & (F469N) & (F547M) & \\
\hline

75 &  1:24:02.35  &  -73:10:01.49  & 19.65 & 19.23 & 18.86 & 18.94 & 18.46 & \\ 
76 &  1:24:27.49  &  -73:10:31.19  & 22.46 & 20.12 & 19.13 & 18.96 & 18.18 & \\ 
77 &  1:24:29.48  &  -73:10:17.94  & 16.24 & 17.70 & 17.82 & 17.94 & 18.05 & \\
78 &  1:24:12.68  &  -73:09:16.29  & 14.65 & 16.20 & 16.43 & 16.46 & 16.46 &\\ 
79 &  1:24:12.71  &  -73:09:13.91  & 16.92 & 18.14 & 18.32 & 18.42 & 18.43 & \\
 
\hline
\end{tabular}
\end{flushleft}
\noindent\\
$^{\dag}$ The capital letters B to H indicate the associated \h2\,
regions presented in Fig.\,\ref{rgb}b. The spectral types and
identifications of Wilcots (\cite{wil}) for some stars are included
inside parentheses.
\end{table*}

\end{document}